\documentclass[aps,prd,amsmath,two column,amssymb]{revtex4}
\usepackage{amsfonts}
\usepackage{amssymb}
\usepackage{txfonts}
\usepackage{mathbbol}
\usepackage{amsfonts}
\usepackage{mathrsfs}
\usepackage{epsfig,bm,dcolumn}
\usepackage{graphicx}
\usepackage{color}
\usepackage{amsmath}
\usepackage{dcolumn}
\usepackage{overpic}
\usepackage{slashed}
\usepackage{hyperref}

\begin{document}
\title{Nambu--Jona-Lasinio model in a parallel electromagnetic field}

\author{Lingxiao Wang$^1$, Gaoqing Cao$^{2,3}$, Xu-Guang Huang$^{3,4}$and Pengfei Zhuang$^1$}
\affiliation{1 Department of Physics, Tsinghua University and Collaborative Innovation Center of Quantum Matter, Beijing 100084, China.\\
2 School of Physics and Astronomy, Sun Yat-Sen University, Guangzhou 510275, China.\\
3 Physics Department and Center for Particle Physics and Field Theory, Fudan University, Shanghai 200433, China.\\
4 Key Laboratory of Nuclear Physics and Ion-beam Application (MOE), Fudan University, Shanghai 200433, China.}
\date{\today}

\begin{abstract}
We explore the features of the $U_A(1)$ and chiral symmetry breaking of the Nambu--Jona-Lasinio model without the Kobayashi-Maskawa-'t Hooft determinant term in the presence of a parallel electromagnetic field. We show that the electromagnetic chiral anomaly can induce both finite neutral pion condensate and isospin-singlet pseudo-scalar $\eta$ condensate and thus modifies the chiral symmetry breaking pattern. In order to characterize the strength of the $U_A(1)$ symmetry breaking, we evaluate the susceptibility associated with the $U_A(1)$ charge. The result shows that the susceptibility contributed from the chiral anomaly is consistent with the behavior of the corresponding $\eta$ condensate. The spectra of the mesonic excitations are also studied.
\end{abstract}


\maketitle
\section{Introduction}
As is well known, the Lagrangian of the quantum chromodynamics (QCD) for light flavors ($u,d$ quarks) has approximate $SU_A(2)$ chiral symmetry and $U_A(1)$ axial symmetry. However, the $U_A(1)$ symmetry will be violated by the chiral anomaly due to the nontrivial topological configurations of the gluon fields~\cite{tHooft:1976snw,Witten:1979vv,Veneziano:1979ec}. The remained $SU_A(2)$ chiral symmetry is also broken spontaneously in vacuum by chiral condensate $\langle\bar{\psi}\psi\rangle\neq 0$ which gives rise to three (pseudo) Goldstone modes identified as pions.

The chiral symmetry breaking and the $U_A(1)$ symmetry breaking are closely related. For example, it was argued that the order and the critical properties of the chiral phase transition are sensitive to the fate of the $U_A(1)$ symmetry at the chiral critical temperature $T_c$~\cite{Pisarski:1983ms}. But it is still unclear whether the $U_A(1)$ symmetry is effectively restored at and above $T_c$. As The nontrivial gluon-field configurations produce both the chiral anomaly and the topological susceptibility, one can use the latter to quantify the strength of the $U_A(1)$ symmetry breaking in both the quenched and the unquenched cases~\cite{Alles:1996nm,Alles:1998mt}. It was found in both cases that the topological susceptibility always drops down above $T_c$ but the topological charge itself still keeps an obvious deviation from zero, which indicates partial restoration of the $U_A(1)$ symmetry and is consistent with the simulation using the instanton model~\cite{Wantz:2009mi} and other effective model~\cite{Jiang:2015xqz}. Recently, there were also important progresses from the lattice QCD simulations, we refer the readers to the Refs.~\cite{Aoki:2012yj,Bazavov:2012qja,Cossu:2013uua,Aoki:2017paw} for more details.

In addition to the temperature and density, the electromagnetic (EM) field provides another way to explore the features of the chiral symmetry breaking and restoration in quark-gluon matter~\cite{Miransky:2015ava}. This kind of study is important because it is relevant to the environments in the compact stars~\cite{Duncan:1992hi,Olausen:2013bpa}, the heavy-ion collisions~\cite{Skokov:2009qp,Deng:2012pc,Deng:2014uja,Bloczynski:2012en}, and the early universe~\cite{Grasso:2000wj} where very strong magnetic fields can exist. Usually, the presence of the magnetic field enhances the chiral condensate in vacuum which is called the magnetic catalysis effect~\cite{Gusynin:1994re,Cao:2014uva}; but the interplay between the magnetic field and the temperature demonstrates the inverse magnetic catalysis effect near $T_c$, the underlying mechanism of which is still not fully understood~\cite{Bali:2011qj,Bali:2012zg,Bruckmann:2013oba,Chao:2013qpa, Endrodi:2015oba,Mao:2016fha}. The effect of the electric field was found always to restore the chiral symmetry because it tends to break the scalar quark-antiquark pairs~\cite{Babansky:1997zh,Klevansky:1989vi,Cao:2015dya}. Furthermore, various chiral-anomaly induced quantum phenomena are also closely related to the EM field, including the chiral magnetic effect~\cite{Kharzeev:2007jp,Fukushima:2008xe}, the chiral magnetic wave~\cite{Kharzeev:2010gd}, the chiral separation effect~\cite{Son:2004tq,Metlitski:2005pr}, the chiral electric separation effect~\cite{Huang:2013iia,Jiang:2014ura}, the anomalous magnetovorticity coupling~\cite{Hattori:2016njk}, the chiral electrodynamics~\cite{Qiu:2016hzd}, etc; see recent reviews~\cite{Liao:2014ava,Kharzeev:2015znc,Huang:2015oca,Hattori:2016emy}.

In Ref.~\cite{Cao:2015cka}, the effect of EM chiral anomaly (which should not be confused with the chiral anomaly due to gluons) on the chiral symmetry breaking and restoration was investigated in the parallel EM field (i.e., the EM field configuration with parallel electric and magnetic fields) and the isospin-triplet neutral pseudo-scalar $\pi^0$ condensate was found to increase with the second Lorentz invariant $I_2={\bf E\cdot B}$ (without loss of generality, we will assume $I_2\geq0$) and to saturate at a critical $I^c_2$. This finding was also confirmed by the calculations using the Wigner function formalism~\cite{Fang:2016uds,Guo:2017dzf}. As the isospin-singlet neutral pseudo-scalar $\eta$ meson (more precisely, the two-flavor counterpart of the $\eta$ meson) has the same quantum numbers as $\pi^0$ meson except for isospin, one can expect that $\eta$ would also condensate in parallel EM field via the EM chiral anomaly, which would then give rise to a macroscopic $U_A(1)$ current divergence via $m_0\langle \bar{\psi}i\gamma^5\psi\rangle$. The purpose of the present paper is to give a detailed study of the properties of the chiral symmetry breaking and $U_A(1)$ symmetry breaking under a parallel EM field. We will adopt the Nambu--Jona-Lasinio (NJL) model with $U_L(2)\otimes U_R(2)$ symmetric interactions in the following discussions, which allows us to illuminate how the sole parallel EM field breaks the $U_A(1)$ symmetry. The vacuum properties and the mesonic fluctuations will be both investigated. Here, it is proper to mention a recent work~\cite{Ruggieri:2016lrn} which studies the generation of chiral density due to Schwinger mechanism and its feedback to the thermodynamic properties of the NJL model. As the feedback is small for a reasonable relaxation time, especially at lower temperatures, we will not discuss such a effect in out study. Our main focus will be the pseudo-scalar $\eta$ and $\pi_0$ channels which were not discussed in Ref.~\cite{Ruggieri:2016lrn}.

The paper is organized as follows. In Sec.~\ref{sec2}, we develop a formalism to evaluate several neutral condensates in a parallel EM field within the chosen NJL model. Section \ref{sec3} is composed of three parts. We first introduce the topological charge to describe the $ U_A(1)$ symmetry breaking in Sec.\ref{sec31}, then the corresponding susceptibility is evaluated to show the strength of $ U_A(1) $ symmetry breaking in Sec.\ref{sec32}, and pole masses of mesonic excitation modes are shown in Sec.\ref{sec33}. A summary will be given in Sec.\ref{sec4}.

\section{Neutral condensates in parallel electromagnetic field }\label{sec2}
As our aim is to study how the parallel EM field influences the $U_A(1)$ and chiral symmetry, we will adopt a two-flavor NJL model without the Kobayashi-Maskawa-'t Hooft (KMT) determinant term so that the Lagrangian density preserves the $U_L(2)\otimes U_R(2)$ symmetry. The Lagrangian in Euclidean space is given by the following form~\cite{Nambu:1961fr,Nambu:1961tp,Klevansky:1989vi},
\begin{eqnarray}\label{LNJL}
{\cal L}_{\rm NJL}=\bar\psi(i\slashed{D}-m_0)\psi+G[(\bar\psi\tau\psi)^2+(\bar\psi i\gamma^5{\tau}\psi)^2],
\end{eqnarray}
where $ \tau=(\tau_0, {\boldsymbol \tau})$ ($ \tau_0$ is the unit matrix and ${\boldsymbol \tau}$ are  Pauli matrices in flavor space), $\psi=(u,d)^T$ represents the two-flavor quark fields, $m_0$ is the current mass of quarks, and $G$ is the four-fermion coupling constant. The parallel EM field is introduced through the covariant derivative $D_\mu=\partial_\mu+iQA_\mu$ with the vector potential chosen as $A_\mu=(iEz,0,-Bx,0)$ ($E,B\geq 0$) and the quark charge matrix $Q={\rm diag}(2/3,-1/3)e$. Note that the presence of the EM field explicitly breaks the symmetry of ${\cal L}_{\rm NJL}$ to $U_A(1)\otimes U_V(1)$.

In order to explore the ground state in this case, we introduce eight auxiliary boson fields, $\sigma=-2G\bar\psi\psi, {\bf a}=-2G\bar\psi{\boldsymbol \tau}\psi$, ${\eta}=-2G\bar\psi i\gamma^5\psi $ and ${\boldsymbol \pi}=-2G\bar\psi i\gamma^5{\boldsymbol \tau}\psi$, via the Hubbard-Stratonavich transformation. Then, by integrating out the quark degrees of freedom, the action is bosonized to the following form:
\begin{eqnarray}\label{action}
{\cal S}_{\rm NJL} &=&\int{d^4x}{\sigma^2+{\boldsymbol{a}}^2+\eta^2+\boldsymbol{\pi}^2\over 4G}\nonumber\\
&&-{\rm Tr}\ln\bigg[i{\slashed D}-m_0-\sigma-{\boldsymbol \tau}{\bf \cdot a}-i\gamma^5\eta-i\gamma^5{\boldsymbol \tau\cdot}{\boldsymbol \pi}\bigg].
\end{eqnarray}
As the charged condensates are energetically unfavored in the EM field~\cite{Cao:2015xja}, it is enough only to consider the following four neutral condensates: $\langle\sigma\rangle\equiv m-m_0, \langle a_3\rangle \equiv\delta m, \langle \eta\rangle\equiv\eta$ and $\langle\pi^0\rangle\equiv\pi^0$.
The corresponding gap equations can be obtained by minimizing the thermodynamic potential with respect to these order parameters, that is, $\partial\Omega/\partial x=0~(x=m,\delta m,\eta,\pi^0)$, which give the following forms:
\begin{eqnarray}
{m-m_0\over 2G}-{1\over\beta V}\text {Tr}\;{\cal S}_{A}(x)&=&0,\label{gapm}\\
{\delta m\over 2G}-{1\over\beta V}\text {Tr}\;{\cal S}_{A}(x)\tau_3&=&0,\label{gapdm}\\
{\eta\over 2G}-{1\over\beta V}\text {Tr}\;{\cal S}_{A}(x)i\gamma^5&=&0,\label{gapeta}\\
{\pi^0\over 2G}-{1\over\beta V}\text {Tr}\;{\cal S}_{A}(x)i\gamma^5\tau_3&=&0.\label{gappi0}
\end{eqnarray}
Here the fermion propagator in the constant EM field is defined as ${\cal S}_{A}(x)=-\left[i{\slashed D}-m-\delta m\tau_3-i\gamma^5\eta-i\gamma^5\pi^0\tau_3\right]^{-1}$
which is diagonal in flavor space. For brevity and convenience which will be illuminated later, we define $m_{u/d}=m\pm\delta m$, $\sigma_{u/d}=\sigma\pm\delta m$, and $\pi^0_{u/d}=\pi^0\pm\eta$. By following the same procedure as in Ref.~\cite{Cao:2015dya}, the propagator of ${\rm f}\ (=u,d)$ favor quark can be written out explicitly in energy-momentum space as
\begin{widetext}
\begin{eqnarray}
	\hat{\cal S}_{\rm f}({p})
	&=&\int_0^\infty {ds}\exp\Big\{-[m_{\rm f}^2+(\pi^0_{\rm f})^2]s-{\tan(q_{\rm f}Es)\over q_{\rm f}E}({p}_4^2+p_3^2)-{\tanh(q_{\rm f}Bs)\over q_{\rm f}B}(p_1^2+p_2^2)\Big\}\nonumber\\
	&&\big[m_{\rm f}\!-\!i~{\rm sgn}(q_{\rm f})\gamma^5\pi^0_{\rm f}-\gamma^4(p_4\!-\!{\tan(q_{\rm f}Es)}p_3)\!-\!\gamma^3(p_3\!+\!{\tan(q_{\rm f}Es)}p_4)\!-\!\gamma^2(p_2\!-\!{i\tanh(q_{\rm f}Bs)}p_1)\nonumber\\
	&&-\gamma^1(p_1+{i\tanh(q_{\rm f}Bs)}p_2)\big]\Big[1-{i\gamma^5\tan(q_{\rm f}Es)\tanh(q_{\rm f}Bs)}
	-i{\gamma^1\gamma^2\tanh(q_{\rm f}Bs)}+{\gamma^4\gamma^3\tan(q_{\rm f}Es)}\Big],\label{propagator}
	\end{eqnarray}
where the Schwinger phase has been dropped because it does not change any conclusion in this work. In order to simplify the discussions, we choose the field strengths as in Ref.~\cite{Cao:2015cka} where the first Lorentz invariant of the EM field $ I_1=B^2-E^2=0$. Substituting this explicit form back to the gap equations Eq.(\ref{gapm}-\ref{gappi0}), we find
	\begin{eqnarray}
	{m-m_0\over 2G}
	&=&N_c\sum_{\rm f=u,d}{1\over4\pi^2}\int_0^\infty {ds\over s^2}e^{-\big(m_{\rm f}^2+(\pi^0_{\rm f})^2\big)s}{m_{\rm f}(q_{\rm f}\sqrt{I_2}s)^2
		\over\tan(q_{\rm f}\sqrt{I_2}s)\tanh(q_{\rm f}\sqrt{I_2}s)}-{N_c\over4\pi^{2}}\sum_{\rm f=u,d}{{\rm sgn}(q_{\rm f})\pi^0_{\rm f}(q_{\rm f}\sqrt{I_2})^2\over{m_{\rm f}^2+(\pi^0_{\rm f})^2}}
	,\label{mgap}\\
	{\delta m\over 2G}&=&N_c\sum_{\rm f=u,d}{1\over4\pi^2}\int_0^\infty {ds\over s^2}e^{-\big(m_{\rm f}^2+(\pi^0_{\rm f})^2\big)s}{{\rm sgn}(q_{\rm f})m_{\rm f}(q_{\rm f}\sqrt{I_2}s)^2
		\over\tan(q_{\rm f}\sqrt{I_2}s)\tanh(q_{\rm f}\sqrt{I_2}s)}-{N_c\over4\pi^{2}}\sum_{\rm f=u,d}{\pi^0_{\rm f}(q_{\rm f}\sqrt{I_2})^2\over{m_{\rm f}^2+(\pi^0_{\rm f})^2}}
	,\label{dmgap}\\
	{\eta\over 2G}&=&N_c\sum_{\rm f=u,d}{1\over4\pi^2}\int_0^\infty {ds\over s^2}e^{-\big(m_{\rm f}^2+(\pi^0_{\rm f})^2\big)s}{{\rm sgn}(q_{\rm f})\pi^0_{\rm f}(q_{\rm f}\sqrt{I_2}s)^2
		\over\tan(q_{\rm f}\sqrt{I_2}s)\tanh(q_{\rm f}\sqrt{I_2}s)}+{N_c\over4\pi^{2}}\sum_{\rm f=u,d}{m_{\rm f}(q_{\rm f}\sqrt{I_2})^2\over{m_{\rm f}^2+(\pi^0_{\rm f})^2}},\label{etagap}\\
	{\pi^0\over 2G}&=&N_c\sum_{\rm f=u,d}{1\over4\pi^2}\int_0^\infty {ds\over s^2}e^{-\big(m_{\rm f}^2+(\pi^0_{\rm f})^2\big)s}{\pi^0_{\rm f}(q_{\rm f}\sqrt{I_2}s)^2
		\over\tan(q_{\rm f}\sqrt{I_2}s)\tanh(q_{\rm f}\sqrt{I_2}s)}+{N_c\over4\pi^{2}}\sum_{\rm f=u,d}{{\rm sgn}(q_{\rm f})m_{\rm f}(q_{\rm f}\sqrt{I_2})^2\over{m_{\rm f}^2+(\pi^0_{\rm f})^2}},\label{pi0gap}
	\end{eqnarray}
where $I_2=EB$.
It is easy to see that Eq.(\ref{mgap})$\pm$Eq.(\ref{dmgap}) and Eq.(\ref{pi0gap})$\pm$Eq.(\ref{etagap}) split the four coupled gap equations to two independent sets of gap equations for $u$ and $d$ quarks, separately. Thus, the thermodynamic potential can be derived consistently by combining the integration over $m$ of Eq.(\ref{mgap}), the integration over $\delta m$ of Eq.(\ref{dmgap}), the integration over $\eta$ of Eq.(\ref{etagap})  and  the integration over $\pi^0$ of Eq.(\ref{pi0gap}). The result is
	\begin{eqnarray}
	\Omega=\sum_{\rm f=u,d}\Bigg[{(m_{\rm f}-m_0)^2+(\pi^0_{\rm f})^2\over4G}+\!{N_c\over8\pi^2}\int_0^\infty\!\! {ds\over s^3}{e^{-\big(m_{\rm f}^2+(\pi^0_{\rm f})^2\big)s}(q_{\rm f}\sqrt{I_2}s)^2
		\over\tan(q_{\rm f}\sqrt{I_2}s)\tanh(q_{\rm f}\sqrt{I_2}s)}\!-\!{N_c\over4\pi^{2}}\!{\rm sgn}(q_{\rm f})\tan^{-1}\Big({\pi^0_{\rm f}\over m_{\rm f}}\Big)
	(q_{\rm f}\sqrt{I_2})^2\Bigg].
	\end{eqnarray}
Using the same regularization scheme as in Ref.~\cite{Cao:2015dya,Cao:2015cka}, the gap equations and the thermodynamic potential become
	\begin{eqnarray}
	{m_{\rm f}-m_0\over 2G}&=&{N_c m_{\rm f}M_{\rm f}\over2\pi^2}\Bigg[\Lambda\Big({1+{\Lambda^2\over M_{\rm f}^2}}\Big)^{1/2}-M_{\rm f}\ln\Big({\Lambda\over M_{\rm f}}
	+\Big({1+{\Lambda^2\over M_{\rm f}^2}}\Big)^{1/2}\Big)\Bigg]+{N_c\over4\pi^2}\int_0^\infty {ds\over s^2}e^{-M_{\rm f}^2s}m_f\Bigg[{(q_{\rm f}\sqrt{I_2}s)^2
		\over\tan(q_{\rm f}\sqrt{I_2}s)\tanh(q_{\rm f}\sqrt{I_2}s)}-1\Bigg]\nonumber\\
	&&-{N_c\over4\pi^{2}}{{\rm sgn}(q_{\rm f})\pi^0_f\over M_{\rm f}^2}
	(q_{\rm f}\sqrt{I_2})^2, \label{mgap1}\\
	{\pi^0_{\rm f}\over 2G}&=&{N_c \pi^0_{\rm f}M_{\rm f}\over2\pi^2}\Bigg[\Lambda\Big({1+{\Lambda^2\over M_{\rm f}^2}}\Big)^{1/2}-M_{\rm f}\ln\Big({\Lambda\over M_{\rm f}}
	+\Big({1+{\Lambda^2\over M_{\rm f}^2}}\Big)^{1/2}\Big)\Bigg]+{N_c\over4\pi^2}\int_0^\infty {ds\over s^2}e^{-M_{\rm f}^2s}\pi^0_{\rm f}\Bigg[{(q_{\rm f}\sqrt{I_2}s)^2
		\over\tan(q_{\rm f}\sqrt{I_2}s)\tanh(q_{\rm f}\sqrt{I_2}s)}-1\Bigg]\nonumber\\
	&&+{N_c\over4\pi^{2}}{{\rm sgn}(q_{\rm f})m_{\rm f}\over M_{\rm f}^2}(q_{\rm f}\sqrt{I_2})^2, \label{pi0gap1}\\
	\Omega&=&\sum_{\rm f=u,d}\Bigg\{{(m_{\rm f}-m_0)^2+(\pi^0_{\rm f})^2\over4G}-{N_cM_{\rm f}^3\over8\pi^2}\Bigg[\Lambda\Big(1+{2\Lambda^2\over M_{\rm f}^2}\Big)\Big({1+{\Lambda^2\over M_{\rm f}^2}}\Big)^{1/2}-M_{\rm f}\ln\Big({\Lambda\over M_{\rm f}}
	+\Big({1+{\Lambda^2\over M_{\rm f}^2}}\Big)^{1/2}\Big)\Bigg]\nonumber\\
	&&+{N_c\over8\pi^2}\int_0^\infty {ds\over s^3}e^{-M_{\rm f}^2s}\Bigg[{(q_{\rm f}\sqrt{I_2}s)^2
		\over\tan(q_{\rm f}\sqrt{I_2}s)\tanh(q_{\rm f}\sqrt{I_2}s)}-1\Bigg]-{N_c\over4\pi^{2}}{\rm sgn}(q_{\rm f})\tan^{-1}\Big({\pi^0_{\rm f}\over m_{\rm f}}\Big)
	(q_{\rm f}\sqrt{I_2})^2\Bigg\},
	\end{eqnarray}
where $M_{\rm f}=[{m^2_{\rm f}+(\pi^0_{\rm f})^2}]^{1/2}$.
\end{widetext}

\begin{figure}[h]
\vspace{0.3cm}
	\centering
	\includegraphics[width=7cm]{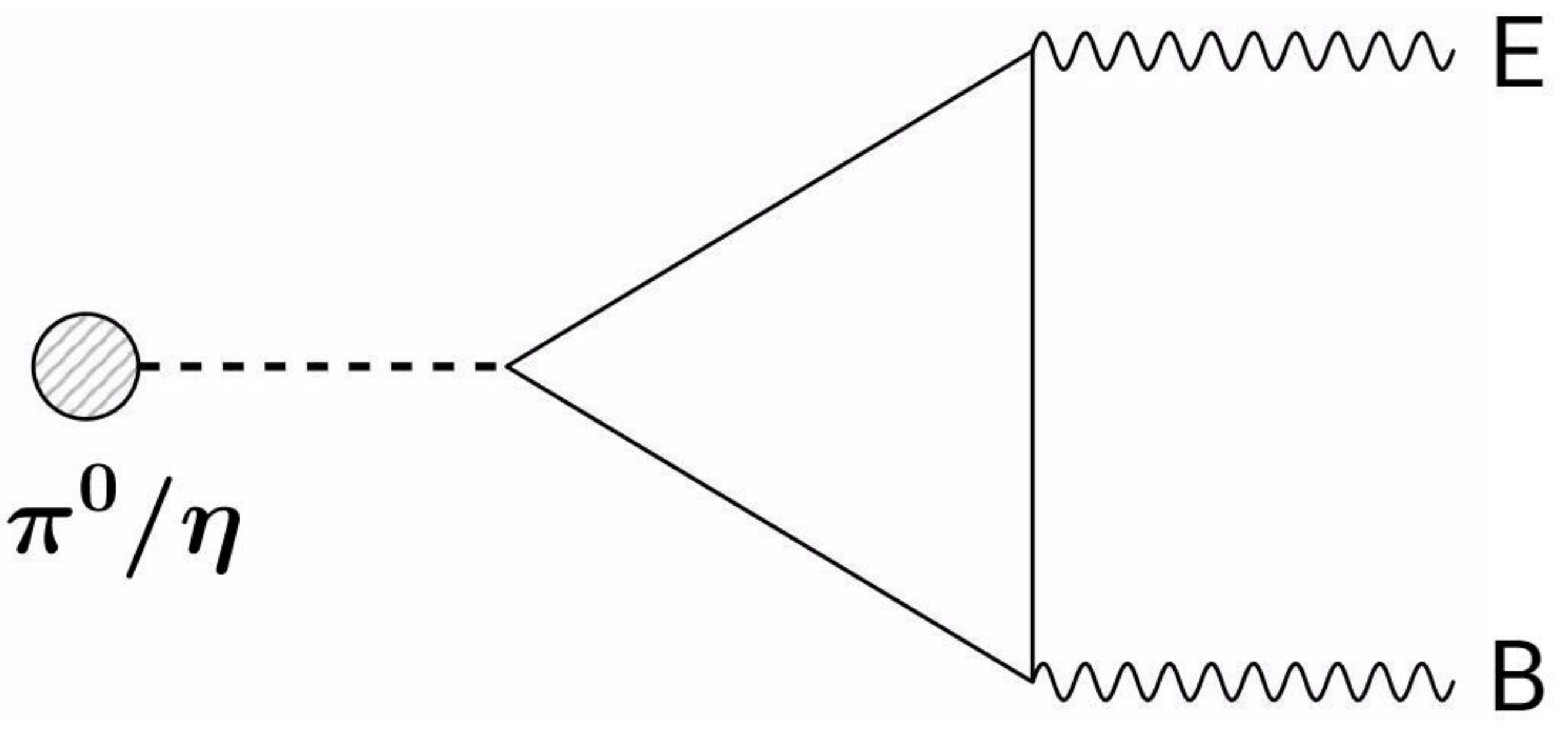}
	\caption{The triangle diagram that is responsible for the $\pi^0$ and $\eta$ condensations in a parallel EM field.}
	\label{feyn}
\end{figure}

It is easy to check from Eq.(\ref{mgap1}) and Eq.(\ref{pi0gap1}) that
\begin{eqnarray}
\pi_{\rm f}^0={\rm sgn}(q_{\rm f}){N_cG\over\pi^2m_0}(q_{\rm f}\sqrt{I_2})^2={\rm sgn}(q_{\rm f}){N_cm^*\over2\pi^{2}m_\pi^2 f_\pi^2}(q_{\rm f}\sqrt{I_2})^2,
\end{eqnarray}
where we have used the Gell-Mann--Oakes--Renner relation in NJL model, $m^2_\pi f_\pi^2=m_0m^*(2G)^{-1}$ with $m^*$ the quark mass in vacuum. Thus, the following model parameter independent results can be extracted,
\begin{eqnarray}
\label{etamf}
\eta&=&{N_cm^*\over4\pi^{2}m_\pi^2 f_\pi^2}\Big[(q_{\rm u}\sqrt{I_2})^2+(q_{\rm d}\sqrt{I_2})^2\Big],\\
\label{pimf}\pi^0&=&{N_cm^*\over4\pi^{2}m_\pi^2 f_\pi^2}\Big[(q_{\rm u}\sqrt{I_2})^2-(q_{\rm d}\sqrt{I_2})^2\Big],
\end{eqnarray}
which can also be obtained from Eqs.(\ref{mgap})-(\ref{etagap}) without explicitly solving them (in the $I_2$ region where $m_{\rm f}>0$). As discussed in Ref.~\cite{Cao:2015cka}, the above $\eta$ and $\pi^0$ condensates are consequence of the EM chiral anomaly as shown in Fig.~\ref{feyn}. Moreover, the above results show that the $\eta$ condensate is always larger than $\pi^0$ condensate under given $I_2$.

\begin{figure}[h]
	\centering
	\includegraphics[width=8cm]{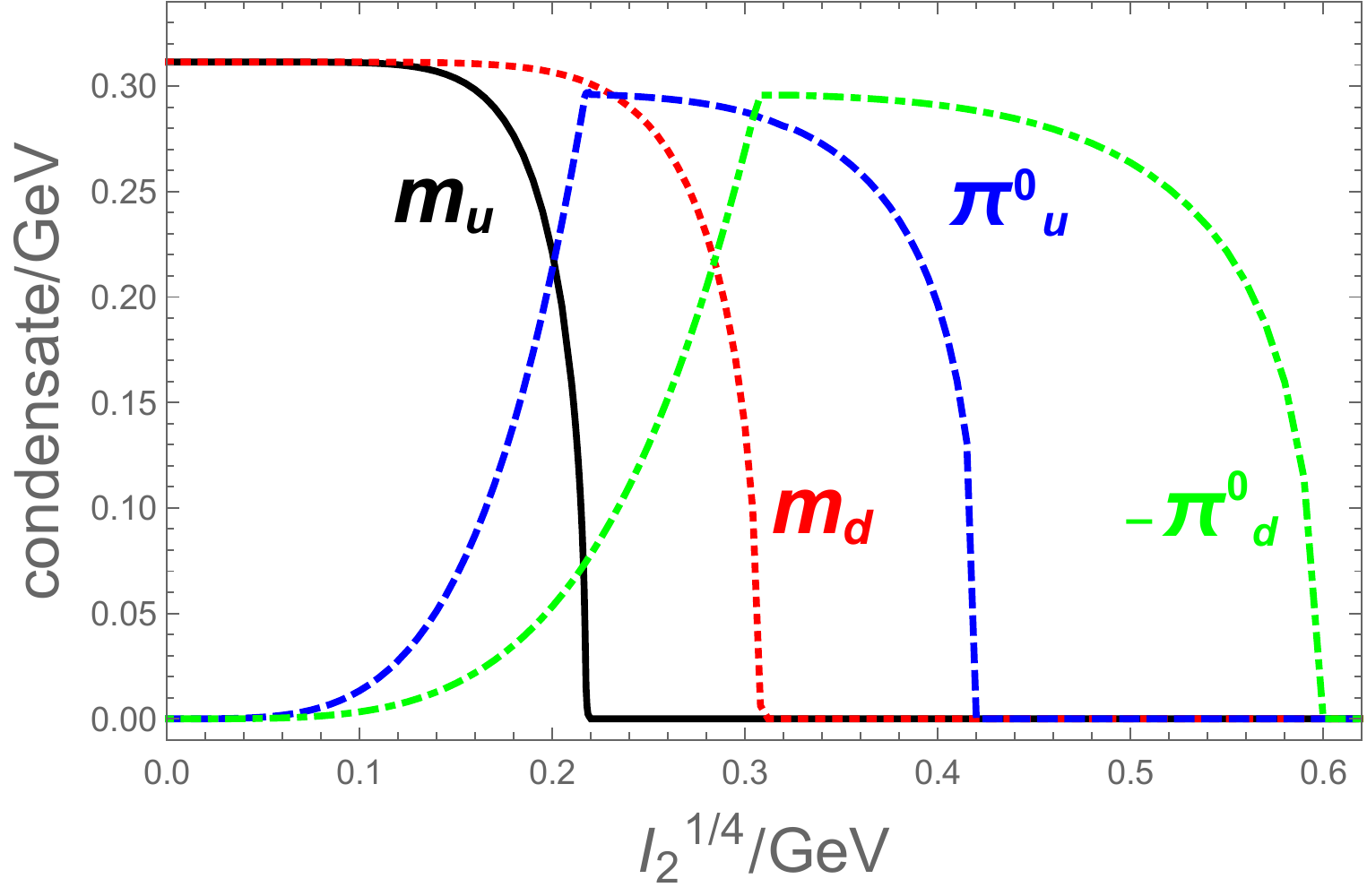}
	\caption{The constitute quark masses $ m_{u/d} $, neutral pion condensates $\pi^0_{\rm u}$ and $-\pi^0_{\rm d}$ as functions of the second Lorentz invariant $I_2$ for the case of $E=B$ in the NJL model.}
	\label{condensate}
\end{figure}

By fixing the model parameters as $G=4.93~{\rm GeV}^{-2}$, $\Lambda=0.653~{\rm GeV}$ and $m_0=5~{\rm MeV}$~\cite{Zhuang:1994dw}, the numerical results by solving the coupled gap equations Eq.(\ref{mgap1}) and Eq.(\ref{pi0gap1}) for $E=B$ are shown in Fig.\ref{condensate}. The solutions to each equation set ${\rm f}=u$ or $d$, i.e., $m_u$ and $\pi^0_u$ or $m_d$ and $\pi^0_d$, show the same features as those found in Ref.~\cite{Cao:2015cka}, though the critical points $I_2^c$ at which $m_{\rm f}=0$ are different for different flavor. The critical $I_2^c$ for $u$ quark corresponds to the peak of $|\delta m|$ while the one for $d$ quark corresponds to the peak of $\eta$. At large enough $I_2$, all the condensates vanish because the strong electric field breaks the mesonic pairs. However, we note that the $U_A(1)$ symmetry is not completely restored at any $I_2$ because the triangle anomaly is always finite in parallel EM field.

As we have stated in the introduction, the electric field and the magnetic field have opposite effects on the chiral condensate at zero temperature. For the case of $E=B$, Fig.\ref{condensate} shows that not only $m_{\rm f}$ but also $M_{\rm f}$ decreases with increasing $I_2$ which indicates that the electric restoration effect takes over the magnetic catalysis effect. This is actually consistent with the previous result in Ref.~\cite{Babansky:1997zh} where similar anti-catalysis effect of $I_2$ on the chiral condensate was observed. To understand this $I_2$ dependence of $M_{\rm f}$, we first note that the decreasing of $m_{\rm f}$ in the region $I_2<I_2^c$ is mainly due to the chiral rotation effect induced by EM chiral anomaly~\cite{Cao:2015cka}, but at the same time the electric field slightly reduces $M_{\rm f}$. For $I_2>I_2^c$, the scalar condensate vanishes, so let us focus on the pseudoscalar condensate $\pi^0_{\rm f}$. In this case (note that $E=B$), the gap equation for $\pi^0_{\rm f}$ can be rewritten as
\begin{widetext}
\begin{eqnarray}
\label{dddd}
{1\over 2G}&=&F(\pi^0_{\rm f})+{N_c\over4\pi^2}\int_0^\infty {ds\over s^2}e^{-(\pi^0_{\rm f})^2s}\Bigg[\Big({q_{\rm f}Es
	\over\tan(q_{\rm f}Es)}-1\Big){q_{\rm f}Bs
	\over\tanh(q_{\rm f}Bs)}+{q_{\rm f}Bs
	\over\tanh(q_{\rm f}Bs)}-1\Bigg],
\end{eqnarray}
with
\begin{eqnarray}
F(\pi^0_{\rm f})&=&N_c{\pi^0_{\rm f}\over2\pi^2}\Bigg[\Lambda\Big({1+{\Lambda^2\over (\pi^0_{\rm f})^2}}\Big)^{1/2}-\pi^0_{\rm f}\ln\Big({\Lambda\over \pi^0_{\rm f}}
+\Big({1+{\Lambda^2\over (\pi^0_{\rm f})^2}}\Big)^{1/2}\Big)\Bigg]
\end{eqnarray}
\end{widetext}
being a monotonically decreasing function of $\pi^0_{\rm f}$. The terms originate from the magnetic field in the integrand of Eq.~(\ref{dddd}) can be reexpressed in terms of the Landau levels by using
\begin{eqnarray}
{1\over\tanh(|q_{\rm f}B|s)}=\sum_{n=0}^\infty(2-\delta_{n0})e^{-2n|q_{\rm f}B|s}.
\end{eqnarray}
For large $I_2$, we can take the lowest Landau level approximation and Eq.~(\ref{dddd}) will just be reduced to the gap equation in a pure electric field. Thus, the behavior of $\pi^0_{\rm f}$ at large $I_2$ would be similar to the case with a pure electric field which favors chiral symmetry restoration~\cite{Klevansky:1989vi}. In a word, it is the Landau levels induced by magnetic field that make the effect of $I_2$ more like electric field.

\section{$U_A(1)$ susceptibility and collective excitations}\label{sec3}
\subsection{Chiral current and $U_A(1)$ charge}\label{sec31}
In the absence of the EM field, the Lagrangian density Eq.(\ref{LNJL}) has the symmetry $ U_A(1)\otimes U_V(1)\otimes SU_L(2)\otimes SU_R(2) $ in chiral limit $ m_0 =0 $. However, in the presence of the EM field, the symmetry $ SU_L(2)\otimes SU_R(2) $ reduces to $ U_L(1)\otimes U_R(1) $ with the neutral transformations $\exp\Big(i{1\pm\gamma^5\over2}\tau_3\theta\Big)$, because $u$ and $d$ quarks have different charges. The $ U_A(1) $ symmetry is still expected to be exact in the chiral limit at classical level, but for finite current quark mass $m_0$ and nonvanishing ${\bf E\cdot B}$, this symmetry is broken with the divergence of chiral current given by
\begin{equation}
\label{diverg}
\partial_\mu J^{\mu}_5=2i\bar{\psi}m_0\gamma^5\psi+N_c{q_{\rm u}^2+q_{\rm d}^2\over2\pi^2}{\bf E\cdot B},
\end{equation}
where $J^{\mu }_5=\bar{\psi}\gamma^\mu\gamma^5\psi$. In equilibrium state, the expectation value $\langle J_5^\mu\rangle$ must be uniform in spacetime and thus the divergence $\partial_\mu\langle J^{\mu}_5\rangle=0$. This gives $\langle\bar{\psi}i\gamma^5\psi\rangle=-N_c({q_{\rm u}^2+q_{\rm d}^2}){\bf E\cdot B}/(4m_0\pi^2)$ which is nothing but Eq.~(\ref{etamf}). Similarly, we can consider the current $J_5^{3\mu}=\bar{\psi}\gamma^\mu\gamma^5 \tau^3\psi$. The vanishing of its divergence at equilibrium can give us the $\pi^0$ condensate which is identical to Eq.~(\ref{pimf}). Therefore, the mean field relations Eq.(\ref{etamf}) and Eq.(\ref{pimf}) have their wholly origins from the EM chiral anomaly at equilibrium with the effective coupling $G$ just giving the definitions of $\eta$ and $\pi^0$.

We now turn to study the fluctuations on top of the mean-field solutions. The first quantity that we want to describe is the $U_A(1)$ susceptibility which characterizes the fluctuation of the $U_A(1)$ charge (which will be defined soon) and the strength of the $U_A(1)$ symmetry breaking. The starting point is the mean-field Lagrangian for quarks
\begin{eqnarray}
	{\cal L}_{\rm MF}=\bar\psi(i\slashed{D}-m-i\gamma^5\eta-\delta m\tau_3-i\gamma^5\tau_3\pi^0)\psi,
\end{eqnarray}
in which the condensates $m, \eta, \delta m, \pi^0$ are all uniform, i.e., they are not dynamical fields. Then one can derive the divergence of the axial current which reads
\begin{equation}\label{anomaly_current}
	\partial_\mu J^{\mu}_5=2i\bar{\psi}m_0\gamma^5\psi+N_c{q_{\rm u}^2+q_{\rm d}^2\over2\pi^2}{\bf E\cdot B}+2N_f Q_A,
\end{equation}
where ($N_f=2$)
\begin{equation}\label{Q}
	Q_A\equiv\frac{1}{N_f}\bar{\psi}[i \gamma^5(\sigma+\delta m \tau_3)-(\eta+\pi^0\tau_3)]\psi.
\end{equation}
One can check that in the ground state specified by the mean-field solutions, the expectation value of $Q_A$ vanishes. In Eq.~(\ref{anomaly_current}), the quantity $Q_A$ is defined in an analogous way as the topological charge in QCD. In QCD, the topological charge represents the violation of the $U_A(1)$ symmetry due to the instanton effects and its susceptibility measures the strength of the violation; see Refs.~\cite{Fukushima:2001hr,Xia:2013caa} for the NJL model mimic of the QCD topological charge. In Eq.~(\ref{anomaly_current}), $Q_A$ represents the violation of the $U_A(1)$ symmetry by the appearance of the condensates $\sigma, \delta m, \pi^0, \eta$ and we will use its susceptibility to quantify the strength of such violation. In this sense, the $Q_A$ term in Eq.~(\ref{anomaly_current}) can be regarded to express a ``spontaneous breaking of $U_A(1)$ symmetry" by the condensates $\sigma, \delta m, \pi^0, \eta$, especially induced by $I_2$. Thus, $Q_A$ has very different meaning from the topological charge and we will therefore call it the $U_A(1)$ charge (should not be confused with the axial charge $J_5^0$).

\subsection{The $U_A(1)$ Susceptibility}\label{sec32}
The topological susceptibility in QCD is a fundamental correlation function and is the key to understand many distinctive dynamics in the $ U_A(1) $ channel. In this section, in order to quantify the strength of the ``spontaneous $ U_A(1) $ symmetry breaking", we
calculate the analogous susceptibility by using the $U_A(1)$ charge density $Q_A$. The $U_A(1)$ susceptibility $ \chi $ can be regarded as the zero energy-momentum limit of the Fourier transformation of the correlation function $  \langle TQ_A(x)Q_A(0)\rangle_C$, that is,
\begin{eqnarray}
\chi&=&\int d^4x \langle TQ_A(x)Q_A(0)\rangle_{C}\nonumber\\
&=&\lim_{k\to0}\int e^{-ikx}d^4x \langle TQ_A(x)Q_A(0)\rangle_{C}\label{ua1_sus}.
\end{eqnarray}
Here, $T$ denotes the time-ordering operator and the subscript $C$ means to pick out only the connected diagrams. Then, by substituting the charge Eq.(\ref{Q}) into Eq.(\ref{ua1_sus}), we can get the explicit form of $ U_A(1) $ susceptibility,
\begin{eqnarray}
\chi&=&\frac{1}{4}\sum_{{\rm f}=u,d}\text{Tr}\int d^4x\langle 0|[\bar{\psi}_{\rm f}(x)(i\gamma^5 \sigma_{\rm f}-{\rm sgn}(q_{\rm f})\pi^0_{\rm f})\psi_{\rm f}(x)\bar{\psi}_{\rm f}(0)\nonumber\\
&&\qquad\qquad\qquad\times(i\gamma^5 \sigma_{\rm f}-{\rm sgn}(q_{\rm f})\pi^0_{\rm f})\psi_{\rm f}(0)]|0\rangle_{C},
\end{eqnarray}
where $\psi_{\rm u}(x)$ and $\psi_{\rm d}(x)$ stand for $u$ and $d$ quark fields, respectively. It is much more convenient to work in energy-momentum space:
\begin{eqnarray}
\chi&=&\chi_1+\chi_2+\chi_3,\nonumber \\
\chi_1&=&-\frac{1}{4}\sum_{\rm f=u,d}(\pi_{\rm f}^0)^2{\rm Tr}\int\frac{d^4p}{(2\pi)^4}\hat{\cal S}_{\rm f}(p)\hat{\cal S}_{\rm f}(p),\label{chi1}\\
\chi_2&=&-\frac{1}{4}\sum_{\rm f=u,d}(\sigma_{\rm f})^2{\rm Tr}\int\frac{d^4p}{(2\pi)^4}\hat{\cal S}_{\rm f}(p)i\gamma^5\hat{\cal S}_{\rm f}(p)i\gamma^5,\\
\chi_3&=&\frac{1}{2}\sum_{\rm f=u,d}{\rm sgn}(q_{\rm f})\sigma_{\rm f}\pi_{\rm f}^0{\rm Tr}\int\frac{d^4p}{(2\pi)^4}\hat{\cal S}_{\rm f}(p)i\gamma^5\hat{\cal S}_{\rm f}(p),
\end{eqnarray}
which are closely related to the mesonic polarization functions as will be shown in the next section.

To evaluate $\chi$ numerically, we need to regularize the above equations as they are divergent. We choose a three-momentum cutoff $\Lambda$ to make the regularization. The regularized susceptibilities, $\chi_i^r$ $i=1,2,3$, can be decomposed in the following way,
\begin{equation}
\chi^r_i(B,E)=[\chi_i(B,E)-\chi_i(0,0)]+\chi^\Lambda_i,
\end{equation}
where the parts in the square bracket are finite and independent of $\Lambda$ and the $\Lambda$ dependent parts, $\chi^\Lambda_i$, are independent of the EM field. Their expressions are
\begin{widetext}
\begin{eqnarray}
	\chi_1(B,E)
	&=&\sum_{\rm f=u,d}\frac{N_c q_{\rm f}^2 I_2}{16\pi^2}(\pi_{\rm f}^0)^2\int_0^\infty sds~{e^{-M_{\rm f}^2 s}}\Bigg[2 {\rm sgn}(q_{\rm f}) m_{\rm f} \pi_{\rm f}^0+\frac{1/s-2m_{\rm f}^2}{\tan(q_{\rm f}\sqrt{I_2}s)\tanh(q_{\rm f}\sqrt{I_2}s)}\Bigg],\\
	\chi_2(B,E)
	&=&\sum_{\rm f=u,d}\frac{N_c q_{\rm f}^2 I_2}{16\pi^2}(\sigma_{\rm f})^2\int_0^\infty  sds~{e^{-M_{\rm f}^2 s}}\Bigg[-2 {\rm sgn}(q_{\rm f}) m_{\rm f} \pi_{\rm f}^0+\frac{1/s-2({\pi_{\rm f}^0})^2}{\tan(q_{\rm f}\sqrt{I_2}s)\tanh(q_{\rm f}\sqrt{I_2}s)}\Bigg],\\
	\chi_3(B,E)
	&=&\sum_{\rm f=u,d}\frac{N_c q_{\rm f}^2 I_2}{8\pi^2}\sigma_{\rm f}\pi_{\rm f}^0\int_0^\infty  sds~{e^{-M_{\rm f}^2 s}}\Bigg[{\rm sgn}(q_{\rm f})(m_{\rm f}^2-{(\pi_{\rm f}^0})^2)+\frac{2m_{\rm f}\pi_{\rm f}^0}{\tan(q_{\rm f}\sqrt{I_2}s)\tanh(q_{\rm f}\sqrt{I_2}s)}\Bigg],
	\end{eqnarray}
\end{widetext}
\begin{eqnarray}
\chi_1^\Lambda&=&\frac{N_c}{4\pi^2}\sum_{\rm f=u,d}(\pi_{\rm f}^0)^2\int_0^\Lambda p^2dp \frac{m_{\rm f}^2+p^2}{(M_{\rm f}^2+p^2)^{3/2}},\\
\chi_2^\Lambda&=&\frac{N_c}{4\pi^2}\sum_{\rm f=u,d}(\sigma_{\rm f})^2\int_0^\Lambda p^2dp \frac{(\pi^0_{\rm f})^2+p^2}{(M_{\rm f}^2+p^2)^{3/2}},\\
\chi_3^\Lambda&=&\frac{N_c}{2\pi^2}\sum_{\rm f=u,d}\int_0^\Lambda p^2dp \frac{m_{\rm f}\sigma_{\rm f}(\pi^0_{\rm f})^2}{(M_{\rm f}^2+p^2)^{3/2}},
\end{eqnarray}

\begin{figure}[h]
	\centering
	\includegraphics[width=8cm]{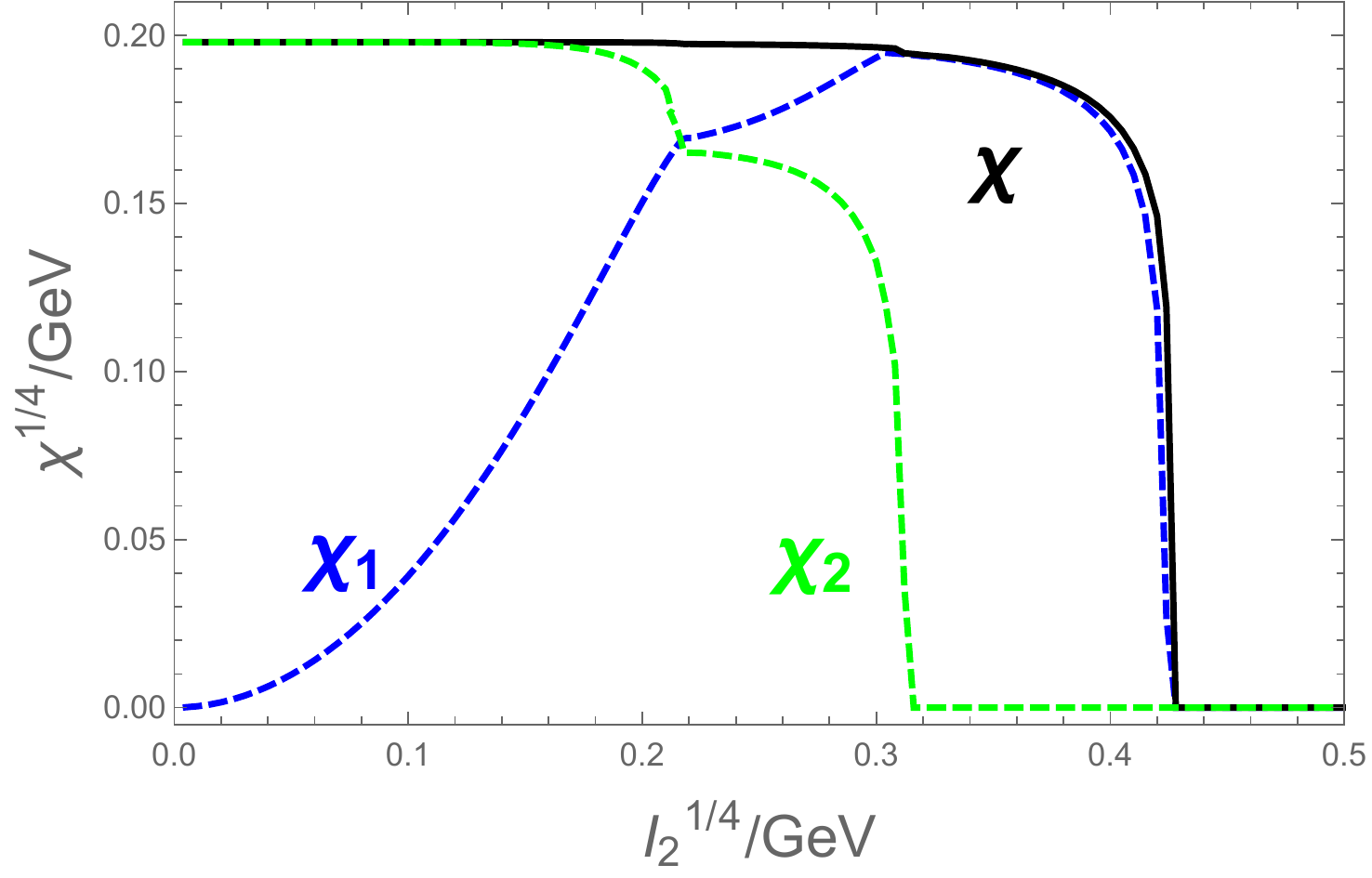}
	\caption[$U_A(1)$ Susceptibility]{ The $U_A(1)$ susceptibility $ \chi $ and its constitute parts $ \chi_1 $ and $\chi_2$ as functions of the second Lorentz invariant $I_2$ for $E=B$ in the NJL model.}
	\label{fig:cchi}
\end{figure}

The numerical results for the total $ U_A(1) $ susceptibility $ \chi $, the one induced by $ \pi^0_{\rm f} $ condensations, $ \chi_1 $, and the one generated by $ \sigma_{\rm f}$ condensations,  $ \chi_2 $, are illuminated in Fig.\ref{fig:cchi} for the case $E=B$. As we can see, $\chi_2$ always decreases with increasing $I_2$ due to both the chiral rotation in $\sigma-\pi^0$ plane and the tendency of chiral restoration as we have analyzed in last section; but $\chi_1$ increases with $I_2$ for not too large EM field because of the chiral rotation~\cite{Cao:2015cka}. The total susceptibility $ \chi $ always decreases with $I_2$. Thus, with increasing $I_2$ not only the chiral symmetry tends to be restored, the $U_A(1)$ symmetry is also effectively restored in terms of $\chi$.

\subsection{Collective excitations}\label{sec33}
We now consider the collective mesonic excitations on top of the mean-field ground state. Comparing to the case of pion superfluidity at large isospin chemical potential or color superconductivity at large baryon chemical potential, the parallel EM field will develop mixing among the flavor collective modes $ \hat{\sigma}, \hat{a}_0, \hat{\pi}_0 $ and $ \hat{\eta} $ in the neutral sector rather than in the charged $ \hat{\pi}^{\pm} $ or color diquark sectors. In the following, we are going to calculate the pole masses of the eigen collective charge-neutral excitations. For the charged modes such as $ \hat{\pi}^{\pm}, \hat{a}^{\pm} $, the masses can be evaluated similarly by neglecting the overall Schwinger phases. However, these charged modes are not the focus of the present paper and we thus will not consider them. Expanding the action Eq.(\ref{action}) to quadratic order of the fluctuation fields which is known as the random phase approximation (RPA), the polarization functions in the neutral sector can be generally written as
\begin{eqnarray}
\Pi_{MM^*}(q)=\int\frac{d^4p}{(2\pi)^4}{\rm Tr}~\hat{S}_{\rm f}(q+p)\Gamma_M\hat{S}_{\rm f}(p)\Gamma_{{M^*}},
\end{eqnarray}
where the interaction vertices are given by
\begin{eqnarray}
\Gamma_M=\Gamma_{M^*}=\left\{\begin{array}{cc}
I,& M=\hat{\sigma}\,; \\
\tau_3,& \,M=\hat{a}_0;\\
i\gamma_5,& \, M=\hat{\eta}\,\, ;\\
i\gamma_5\tau_3,& \ \ M=\hat{\pi}_0.\end{array}\right.
\end{eqnarray}

\begin{figure}[h]
	\includegraphics[width=9cm]{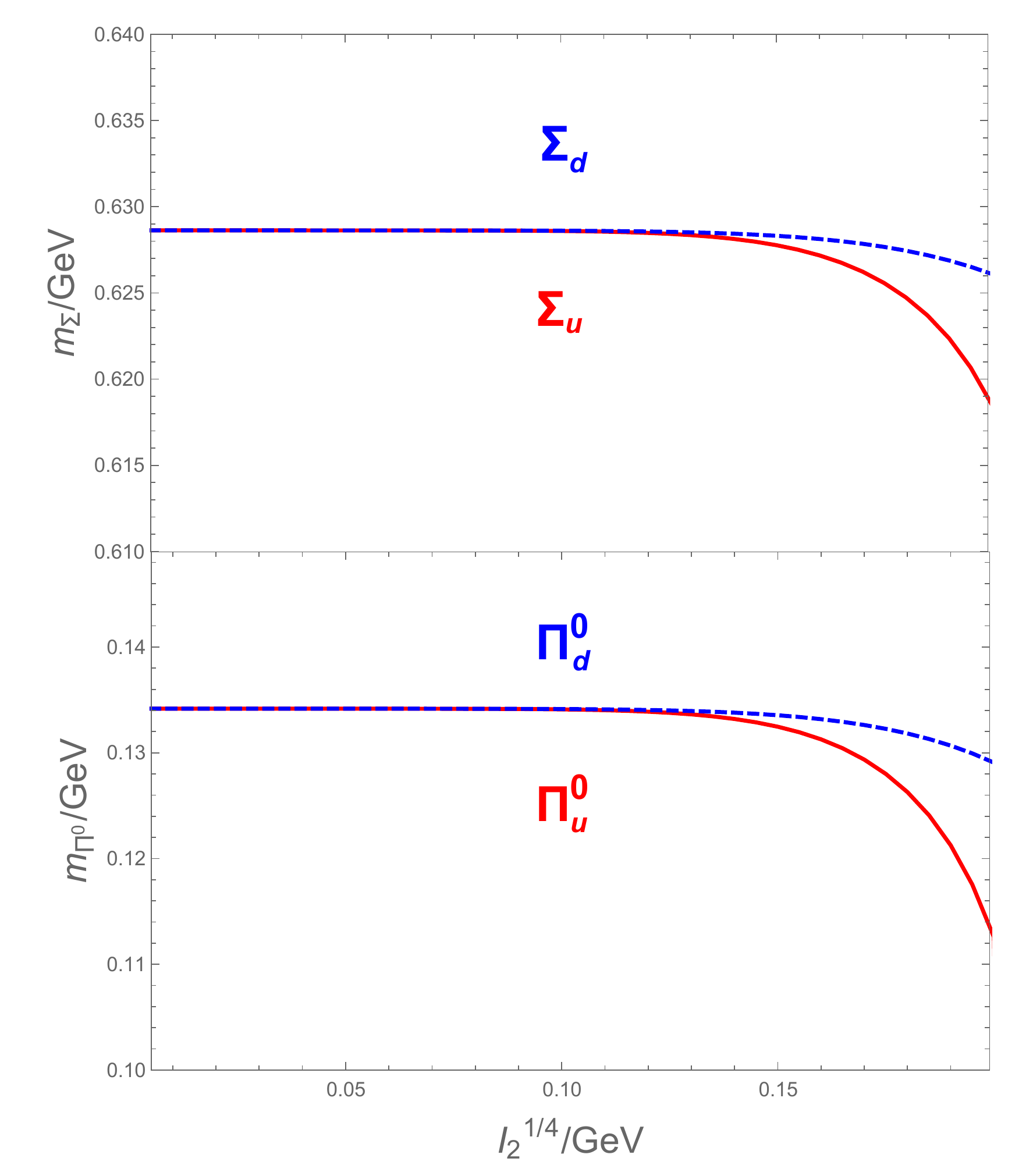}
	\caption[collective modes]{ The pole masses of the eigen collective excitations as functions of $ I_2 $ in the case of $E=B$. The upper panel shows the $ \Sigma $ sector and the lower one shows the $ \Pi $ sector. }
	\label{fig:mesonspectrum}
\end{figure}

The pole masses of the collective excitations are obtained by setting $q_4=im_M$ and $q_{i\neq4}=0 $ in the corresponding Euclidean propagators. We preserve the lengthy derivations of the polarization functions in Appendix.~\ref{polarization}. In the matrix form, the polarization function can be represented as the following:
\begin{eqnarray}
\Pi(q_4)=\left(\begin{array}{cccc}
\Pi_{\hat{\sigma}\hat{\sigma}}(q_4) & \Pi_{\hat{\sigma}\hat{a}_0}(q_4) & \Pi_{\hat{\sigma}\hat{\eta}}(q_4) & \Pi_{\hat{\sigma}\hat{\pi}_0}(q_4) \\
\Pi_{\hat{a}_0\hat{\sigma}}(q_4) & \Pi_{\hat{a}_0\hat{a}_0}(q_4) & \Pi_{\hat{a}_0\hat{\eta}}(q_4) & \Pi_{\hat{a}_0\hat{\pi}_0}(q_4) \\
\Pi_{\hat{\eta}\hat{\sigma}}(q_4) & \Pi_{\hat{\eta}\hat{a}_0}(q_4) & \Pi_{\hat{\eta}\hat{\eta}}(q_4) & \Pi_{\hat{\eta}\hat{\pi}_0}(q_4) \\
\Pi_{\hat{\pi}_0\hat{\sigma}}(q_4) & \Pi_{\hat{\pi}_0\hat{a}_0}(q_4) & \Pi_{\hat{\pi}_0\hat{\eta}}(q_4) & \Pi_{\hat{\pi}_0\hat{\pi}_0}(q_4)
\end{array} \right).
\end{eqnarray}
Then, the inverse of the effective mesonic propagator in the matrix form is given by
\begin{eqnarray}
\mathcal{G}^{-1}(q_4)=\frac{1}{2G}-\Pi(q_4).
\end{eqnarray}
By diagonalizing $\mathcal{G}^{-1} $, we obtain the inverse propagator of the mass eigen modes which we denote as $\Sigma_{\rm u}, \Sigma_{\rm d}, \Pi^0_{\rm u}$, and $\Pi_{\rm d}^0$:
\begin{eqnarray}
\mathcal{G}^{-1}_{\Sigma_{\rm u}}=\frac{1}{2G}\!+\!\Pi^r_{\hat{\sigma}_{\rm u}\hat{\sigma}_{\rm u}}\!+\!\Pi^r_{\hat{\pi}^0_{\rm u}\hat{\pi}^0_{\rm u}}\!+\!\!\sqrt{(\Pi^r_{\hat{\sigma}_{\rm u}\hat{\sigma}_{\rm u}}\!\!-\!\Pi^r_{\hat{\pi}^0_{\rm u}\hat{\pi}^0_{\rm u}})^2\!+\!4(\Pi^r_{\hat{\sigma_{\rm u}}\hat{\pi}^0_{\rm u}})^2} ,\nonumber\\
\mathcal{G}^{-1}_{\Pi^0_{\rm u}}=\frac{1}{2G}\!+\!\Pi^r_{\hat{\sigma}_{\rm u}\hat{\sigma}_{\rm u}}\!+\!\Pi^r_{\hat{\pi}^0_{\rm u}\hat{\pi}^0_{\rm u}}\!-\!\!\sqrt{(\Pi^r_{\hat{\sigma}_{\rm u}\hat{\sigma}_{\rm u}}\!\!-\!\Pi^r_{\hat{\pi}^0_{\rm u}\hat{\pi}^0_{\rm u}})^2\!+\!4(\Pi^r_{\hat{\sigma_{\rm u}}\hat{\pi}^0_{\rm u}})^2} ,\nonumber\\
\mathcal{G}^{-1}_{\Sigma_{\rm d}}=\frac{1}{2G}\!+\!\Pi^r_{\hat{\sigma}_{\rm d}\hat{\sigma}_{\rm d}}\!+\!\Pi^r_{\hat{\pi}^0_{\rm d}\hat{\pi}^0_{\rm d}}\!+\!\!\sqrt{(\Pi^r_{\hat{\sigma}_{\rm d}\hat{\sigma}_{\rm d}}\!\!-\!\Pi^r_{\hat{\pi}^0_{\rm d}\hat{\pi}^0_{\rm d}})^2\!+\!4(\Pi^r_{\hat{\sigma_{\rm d}}\hat{\pi}^0_{\rm d}})^2} ,\nonumber\\
\mathcal{G}^{-1}_{\Pi^0_{\rm d}}=\frac{1}{2G}\!+\!\Pi^r_{\hat{\sigma}_{\rm d}\hat{\sigma}_{\rm d}}\!+\!\Pi^r_{\hat{\pi}^0_{\rm d}\hat{\pi}^0_{\rm d}}\!-\!\!\sqrt{(\Pi^r_{\hat{\sigma}_{\rm d}\hat{\sigma}_{\rm d}}\!\!-\!\Pi^r_{\hat{\pi}^0_{\rm d}\hat{\pi}^0_{\rm d}})^2\!+\!4(\Pi^r_{\hat{\sigma_{\rm d}}\hat{\pi}^0_{\rm d}})^2},
\end{eqnarray}
where $\hat{\sigma}_{\rm u/d}=(\hat{\sigma}\pm\hat{a})/2$ and $\hat{\pi}^0_{\rm u/d}=(\hat{\pi}^0\pm\hat{\eta})/2$.
One should notice that these eigen modes are linear combinations of the original modes $ \sigma, a_0, \eta, \pi^0 $ and are diagonal in flavor space. The fields $\Sigma_{\rm u}$ and $\Sigma_{\rm d}$ are dominated by the $\sigma$ sector while the fields $\Pi^0_{\rm u}$ and $\Pi^0_{\rm d}$ are dominated by the $\pi^0$ sector for small $I_2$, but the dominations exchange around the end of chiral rotation. Finally, the pole masses of these modes can be obtained by setting $\mathcal{G}^{-1}_M(q_4=im_M)=0~(M=\Sigma_{\rm u}, \Sigma_{\rm d}, \Pi^0_{\rm u}, \Pi_{\rm d}^0)$ and the numerical results are shown in Fig.\ref{fig:mesonspectrum}.

For clarity, we choose the range of $ I_2^{1/4} $ from $0$ to $0.2 GeV$ where $ U_A(1) $ susceptibility is the largest. In this region, all the pole masses of collective excitations evolve slowly with $I_2$, but the masses of $\Sigma_{\rm u}$ and $\Pi^0_{\rm u}$ drop a bit faster than those of $\Sigma_{\rm d}$ and $\Pi^0_{\rm d}$ due to the larger charge of $u$ quark than that of $d$ quark. To understand the decreasing features with respect to $I_2$, we can consider the small $I_2$ limit. In this case, the masses of the lighter modes $\Pi^0_{\rm u}$ and $\Pi^0_{\rm d}$ are
\begin{eqnarray}
m_{\Pi^0_{\rm u}}&\approx& m_\pi\left(1-\frac{N_c}{48\pi^2}\frac{g^2_{\pi qq}}{m^{*4}}q_u^2 I_2\right),\nonumber\\
m_{\Pi^0_{\rm d}}&\approx& m_\pi\left(1-\frac{N_c}{48\pi^2}\frac{g^2_{\pi qq}}{m^{*4}}q_d^2 I_2\right),
\end{eqnarray}
where the slops are negative and proportional to $q_{\rm f}^2$, which are then qualitatively consistent with the numerical results.

\section{Summary}\label{sec4}
In this paper, we study the NJL model with a $U_L(2)\otimes U_R(2)$ symmetry under a parallel EM field at zero temperature and quark chemical potential. In particular, we focus on the breaking and restoration of the chiral symmetry and the $U_A(1)$ symmetry as the second Lorentz invariant $I_2$ varies. This study extends the previous work~\cite{Cao:2015cka} to include both isospin singlet $\eta$ condensation and mass splitting $\delta m$ between different flavors. In such a way, the four coupled gap equations Eq.(\ref{mgap}-\ref{pi0gap}) can be split into two independent equation sets for $u$ and $d$ quarks, respectively.

Our first finding is that the EM chiral anomaly induces not only the isospin triplet $\pi^0$ condensation but also the isospin singlet $\eta$ condensation.
The result is presented in Fig.~\ref{condensate}. Although the quark masses $m_{\rm f}$ and neutral pion condensate $\pi^0$ show quite similar features as the previous work~\cite{Cao:2015cka}, the $\eta$ and mass splitting $\delta m$ behave differently. Actually, the critical $I_2$ for $u$ and $d$ quarks correspond to the peaks of $\delta m$ and $\eta$ condensates, respectively. In order to show the strength of $U_A(1)$ symmetry breaking in the parallel EM field, we calculate the susceptibility by defining a $U_A(1)$ charge in analogue to the topological charge in QCD. The total susceptibility $\chi$ decreases with $I_2$ indicating an effective decrement of the $U_A(1)$ symmetry breaking. At last, we explore the eigen excitation modes, the pole masses of which all decrease with $I_2$ as shown in Fig.~\ref{fig:mesonspectrum}.

Finally, we comment about the stability of the $\pi^0$ and $\eta$ condensed vacuum. Under the exertion of the electric field, the charged particle-antiparticle pairs (mostly $\pi^\pm$ in the confined phase) can be induced through the Schwinger mechanism and may drive the vacuum unstable. However, as discussed in Ref.~\cite{Cao:2015cka}, for a parallel EM field with the configuration of $E=B$, such a pair production rate is strongly suppressed due to the enhancement of the charged pion mass by the magnetic field. Therefore, we are eligible to consider the ``equilibrium" property of the vacuum.

Most recently, this work has been extended to the case with finite temperature and quark chemical potential~\cite{Wang:2017pje}. In the future, this work can also be extended to three-flavor NJL model with the KMT determinant. Then the effect of the interplay between QCD anomaly represented by KMT determinant and QED anomaly induced by parallel EM field on both the ground state and meson properties can be studied.  

\emph{Acknowledgments}---
LW and PZ are supported by the NSFC and MOST grant No. 11335005, 11575093, 2013CB922000 and 2014CB845400. GC and XGH are supported by the Thousand Young Talents Program of China and NSFC with Grant No.~11535012 and No.~11675041. GC is also supported by China Postdoctoral Science Foundation with Grant No. KLH1512072.

\appendix
\begin{widetext}
\section{Polarization Functions}\label{polarization}
In this appendix, we derive the polarization functions involved in the neutral sector by adopting the imaginary proper time presentations for the quark propagators~\cite{Klevansky:1989vi} and finally regularize them as in Ref \cite{Cao:2015cka}. In this way, the proper time integrations are well defined and the pole masses of collective excitations can be estimated numerically. The polarization function of $ \pi^0 $ with transformation energy $ q_4 $ nonzero can be evaluated as the following,
\begin{eqnarray}
&&\Pi_{\hat{\pi}^0\hat{\pi}^0}({B},{E},q_4)\equiv\int\frac{d^4 p}{(2\pi)^4}{\rm Tr}   \hat{S}(p+q_4)i\gamma_5\tau_3\hat{S}(p)i\gamma_5\tau_3\nonumber\\
&=&-N_c\sum_{\rm f=u,d}\int_{0}^{\infty}ds\int_{0}^{\infty}ds'\int\frac{d^4 p}{(2\pi)^4}~{\rm exp}\Bigg\{ i \Bigg[-M_{\rm f}^2(s+s')-(\frac{{\rm tan}(q_{\rm f} B s)}{q_{\rm f} B}+\frac{{\rm tan}(q_{\rm f} B s')}{q_{\rm f} B})(p_2^2+p_1^2)-\frac{\tanh(q_{\rm f} E s')}{q_{\rm f} E}(p_4^2+p_3^2)-\nonumber\\
&&\frac{\tanh(q_{\rm f} E s)}{q_{\rm f} E}((p_4+q_4)^2+p_3^2)\Bigg]\Bigg\}~{\rm tr}[m_{\rm f}-{\rm sgn}(q_{\rm f})i\gamma_5\pi_{\rm f}^0-\gamma^4((p_4+q_4)-i\tanh(q_{\rm f} E s)p_3)-\gamma^3(p_3+i\tanh(q_{\rm f} Es)(p_4+q_4))\nonumber\\
&&-\gamma^2(p_2+{\rm tan}(q_{\rm f} B s)p_1)-\gamma^1(p_1-{\rm tan}(q_{\rm f} B s)p_2)][1+i\gamma_5\tanh(q_{\rm f} Es){\rm tan}(q_{\rm f} Bs)+\gamma^1\gamma^2{\rm tan}(q_{\rm f} Bs)+i\gamma^4\gamma^3\tanh(q_{\rm f} Es)]i\gamma_5\nonumber\\
&&[m_{\rm f}-{\rm sgn}(q_{\rm f})i\gamma_5\pi_{\rm f}^0-\gamma^4(p_4-i\tanh(q_{\rm f} E s')p_3)-\gamma^3(p_3+i\tanh(q_{\rm f} Es')p_4)-\gamma^2(p_2+{\rm tan}(q_{\rm f} B s')p_1)-\gamma^1(p_1-{\rm tan}(q_{\rm f} B s')p_2)]\nonumber\\
&&[1+i\gamma_5\tanh(q_{\rm f} Es'){\rm tan}(q_{\rm f} Bs')+\gamma^1\gamma^2{\rm tan}(q_{\rm f} Bs')+i\gamma^4\gamma^3\tanh(q_{\rm f} Es')]i\gamma_5\nonumber
\end{eqnarray}
\begin{eqnarray}
&=&N_c\sum_{\rm f=u,d}\frac{q_{\rm f}Eq_{\rm f}B}{8\pi^2}\int_{0}^{\infty}tdt\int_{-1}^{1}du~{\rm exp}\Bigg\{-i [M_{\rm f}^2t+\frac{\tanh(q_{\rm f}E\frac{t(1+u)}{2})\tanh(q_{\rm f} E\frac{t(1-u)}{2})}{q_{\rm f} E(\tanh(q_{\rm f} E\frac{t(1+u)}{2})+\tanh(q_{\rm f} E\frac{t(1-u)}{2}))}q_4^2\Bigg]\Bigg\}\nonumber\\
&&\Bigg[-2{\rm sgn}(q_{\rm f})m_{\rm f}\pi_{\rm f}^0+\frac{q_4^2 \tanh(q_{\rm f} E\frac{t(1+u)}{2})\tanh(q_{\rm f} E\frac{t(1-u)}{2}){\rm sinh^{-2}}(q_{\rm f} Et)}{(\tanh(q_{\rm f} E\frac{t(1+u)}{2})+\tanh(q_{\rm f} E\frac{t(1-u)}{2})){\rm tan}(q_{\rm f} Bt)}+\frac{1}{\tanh(q_{\rm f} Et){\rm tan}(q_{\rm f} Bt)}\Big(\frac{i}{t}+2(\pi_{\rm f}^0)^2-q_4^2\frac{1}{2} \text{csch}(q_{\rm f} Et)\nonumber\\
&& (u \sinh (q_{\rm f} Et u)-\coth (q_{\rm f} Et)\cosh(q_{\rm f} Et u)+\text{csch}(q_{\rm f} Et))\Big)\Bigg],\label{pipi}
\end{eqnarray}
where we've used partial integral to remove ${\rm sin^{-2}}(q_{\rm f} Bt)$ in the last step due to the non-overlapping condition~\cite{Schwinger1973}. Similarly, the $\sigma$-mode polarization function and the corresponding mixing term can be given as the following:
\begin{eqnarray}
&&\Pi_{\hat{\sigma}\hat{\sigma}}({B},{E},q_4)\equiv\int\frac{d^4 p}{(2\pi)^4}{\rm Tr}\hat{S}(p+q_4)\hat{S}(p)\nonumber\\
&=&N_c\sum_{\rm f=u,d}\frac{q_{\rm f}Eq_{\rm f}B}{8\pi^2}\int_{0}^{\infty}tdt\int_{-1}^{1}du~{\rm exp}\Bigg\{-i \Bigg[M_{\rm f}^2t+\frac{\tanh(q_{\rm f}E\frac{t(1+u)}{2})\tanh(q_{\rm f} E\frac{t(1-u)}{2})}{q_{\rm f} E(\tanh(q_{\rm f} E\frac{t(1+u)}{2})+\tanh(q_{\rm f} E\frac{t(1-u)}{2}))}q_4^2\Bigg]\Bigg\}\nonumber\\
&&\Bigg[2{\rm sgn}(q_{\rm f})m_{\rm f}\pi_{\rm f}^0+\frac{q_4^2 \tanh(q_{\rm f} E\frac{t(1+u)}{2})\tanh(q_{\rm f} E\frac{t(1-u)}{2}){\rm sinh^{-2}}(q_{\rm f} Et)}{(\tanh(q_{\rm f} E\frac{t(1+u)}{2})+\tanh(q_{\rm f} E\frac{t(1-u)}{2})){\rm tan}(q_{\rm f} Bt)}+\frac{1}{\tanh(q_{\rm f} Et){\rm tan}(q_{\rm f} Bt)}\Big(\frac{i}{t}+2m_{\rm f}^2-q_4^2\frac{1}{2} \text{csch}(q_{\rm f} Et) \nonumber\\
&&(u \sinh (q_{\rm f} Et u)-\coth (q_{\rm f} Et)\cosh(q_{\rm f} Et u)+\text{csch}(q_{\rm f} Et))\Big)\Bigg],\label{sisi}\\
&&\Pi_{\hat{\sigma}\hat{\pi}^0}({B},{E},q_4)=\Pi_{\hat{\pi}^0\hat{\sigma}}({B},{E},q_4)\equiv\int\frac{d^4 p}{(2\pi)^4}{\rm Tr}   \hat{S}(p+q_4)i\gamma_5\tau_3\hat{S}(p)\nonumber\\
&=&-N_c\sum_{\rm f=u,d}\frac{q_{\rm f}Eq_{\rm f}B}{8\pi^2}\int_{0}^{\infty}tdt\int_{-1}^{1}du~{\rm exp}\Bigg\{-i\Bigg[M_{\rm f}^2t+\frac{\tanh(q_{\rm f}E\frac{t(1+u)}{2})\tanh(q_{\rm f} E\frac{t(1+u)}{2})}{q_{\rm f} E(\tanh(q_{\rm f} E\frac{t(1+u)}{2})+\tanh(q_{\rm f} E\frac{t(1+u)}{2}))}q_4^2\Bigg]\Bigg\}\nonumber\\
&&\Bigg[m_{\rm f}^2-(\pi_{\rm f}^0)^2+\frac{2{\rm sgn}(q_{\rm f})m_{\rm f}\pi_{\rm f}^0}{{\rm tan}(q_{\rm f} Bt)\tanh(q_{\rm f} Et)}\Bigg],\label{sipi}
\end{eqnarray}
and the other polarization functions can be easily obtained by modifying the three equations Eq.(\ref{pipi}-\ref{sipi}).
\end{widetext}
Then, by following the "vacuum regularization" scheme as in Ref.~\cite{Cao:2015dya,Cao:2015cka}, the regularized forms of the polarization functions can be written as
\begin{eqnarray}
	\Pi_{MM^*}^r(q_4)&=&[\Pi_{MM^*}(B,E,q_4)-\lim_{B,E\rightarrow0}\Pi_{MM^*}(B,E,q_4)]\nonumber\\
	&&+\Pi_{MM^*}^\Lambda(q_4),
\end{eqnarray}
where $ \Pi_{MM^*}^\Lambda(q_4) $ are the polarization functions with vanishing EM field which can be regularized by three momentum cutoff $ \Lambda $ as
\begin{widetext}
\begin{eqnarray}
\Pi^{\Lambda}_{\hat{\pi}^0\hat{\pi}^0}&=&N_c\int\frac{d^4 p}{(2\pi)^4}\Bigg[\frac{8 \left(m_{\rm f}^2-(\pi_{\rm f}^0)^2+p_1^2+p_2^2+p_3^2+p_4 (p_4+q_4)\right)}{\left(m_{\rm f}^2+(\pi_{\rm f}^0)^2+p_1^2+p_2^2+p_3^2+p_4^2\right) \left(m_{\rm f}^2+(\pi_{\rm f}^0)^2+p_1^2+p_2^2+p_3^2+(p_4+q_4)^2\right)}\Bigg]\nonumber\\
&=&N_c\int_0^\Lambda p^2dp\frac{8 \left(m_{\rm f}^2+p^2\right)}{\pi ^2 \sqrt{m_{\rm f}^2+p^2+(\pi_{\rm f}^0)^2} \left(4 m_{\rm f}^2+4 p^2+4(\pi_{\rm f}^0)^2+q_4^2\right)},\\
\Pi^{\Lambda}_{\hat{\sigma}\hat{\sigma}}&=&N_c\int_0^\Lambda p^2dp\frac{8 \left((\pi_{\rm f}^0)^2+p^2\right)}{\pi ^2 \sqrt{m_{\rm f}^2+p^2+(\pi_{\rm f}^0)^2} \left(4 m_{\rm f}^2+4 p^2+4(\pi_{\rm f}^0)^2+q_4^2\right)},\\
\Pi^{\Lambda}_{\hat{\pi}^0\hat{\sigma}}&=&-N_c\int_0^\Lambda p^2dp\frac{8\pi_{\rm f}^0 m_{\rm f}}{\pi ^2 \sqrt{m_{\rm f}^2+p^2+(\pi_{\rm f}^0)^2} \left(4 m_{\rm f}^2+4 p^2+4(\pi_{\rm f}^0)^2+q_4^2\right)}.
\end{eqnarray}
\end{widetext}

\end{document}